\documentclass[aps,prb,twocolumn,showpacs]{revtex4}

\usepackage{amsmath}
\usepackage{epsfig}
\newcommand{\bea}{\begin{eqnarray}}
\newcommand{\eea}{\end{eqnarray}}
\newcommand{\be}{\begin{equation}}
\newcommand{\ee}{\end{equation}}

\begin{document}

\title{Transport in a spin incoherent Luttinger Liquid}

\author{Gregory A. Fiete,$^1$ Karyn Le Hur,$^2$ and Leon Balents$^3$}

\affiliation{$^1$Kavli Institute for Theoretical Physics, University of California, Santa Barbara, CA 93106, USA\\$^2$D\'epartment de Physique and RQMP, Universit\'e De Sherbrooke, Sherbrooke, Qu\'ebec, Canada, J1K 2R1
\\$^3$Physics Department, University of California, Santa Barbara, CA 93106, USA}

\begin{abstract}

We theoretically investigate transport in a spin incoherent one
dimensional electron system, which may be
realized in quantum wires at low electron density and finite
temperature.   Both the pure and disordered cases are considered, both
in finite wires and in the thermodynamic limit.  The effect of Fermi
liquid leads attached to the finite length system is also
addressed. In the infinite system, we find a phase diagram identical to
that obtained for a spinless Luttinger Liquid, provided we make the
identification $g=2g_c$, where $g$ is the interaction parameter in a
spinless Luttinger Liquid and $g_c$ is the interaction parameter of
the charge sector of a Luttinger Liquid theory for electrons with
spin. For a finite length wire attached to Fermi liquid leads the
transport depends on details of the disorder in the wire. A simple
picture for the cross-over from the spin incoherent regime to the spin
coherent regime as the temperature is varied is also discussed, as
well as some physical implications.

\end{abstract}

\date{\today}
%\pacs{74.21.-b,71.Pm,73.21.Hb,73.23.Hk}
\maketitle

%73.21.-b Electron states and collective excitations in multilayers, quantum
%wells, mesoscopic, and nanoscale systems
%71.10.Pm Fermions in reduced dimensions
%73.21.Hb Quantum Wires
%73.23.Hk Coulomb blockade; single-electron tunneling

%71.10.Pm Fermions in reduced dimensions
%71.27.+a Strongly correlated electron systems;heavy fermions
%73.21.-b Electron states and collective excitations in multilayers, quantum
%wells, mesoscopic, and nanoscale systems

\section{Introduction}
\label{sec:intro}

Low dimensional electron systems have attracted much attention in
recent years because they provide an opportunity to realize
exceptionally rich physics not readily found in higher dimensions.
One such example is the so called Luttinger-Liquid (LL) state of an
electron gas.\cite{haldane81,Voit:rpp95} The LL state is characterized
by gapless excitations, and in the case of an electron gas,
spin-charge separation realized by separate collective spin and charge
excitations, each with its distinct propagation velocity.  The
tunneling density of states also exhibits a characteristic power law
suppression at low energies as a result of an ``orthogonality
catastrophe'' at zero energy from the re-arranging of the
wavefunctions of the electrons to accommodate the new (tunneling)
particle. Experimental evidence for the LL state in one dimensional
electron systems is by now irrefutable, with measurements showing both
the characteristic power law suppression of the tunneling density of
states, and measurements of the spectral function providing direct
evidence of spin-charge separation, including quantitative measures of
the respective collective mode
velocities. \cite{Steinberg:sci05,ishii:nat03,Bockrath:nat99,Yao:nat99}

Real samples measured in experiments are not perfectly clean, and it
is important to understand the effects of impurities on the electronic
properties, such as transport, in one-dimensional (1-d) interacting
electron systems.  Pioneering work by Kane and
Fisher,\cite{Kane:prl92,Kane:prb92} and by Furusaki and
Nagaosa\cite{Furusaki:prb93,Nagaosa:prb93} established the central
results for transport in a single and double impurity system in a LL,
both at weak and strong impurity strengths. Since then, beautiful
numerical studies have confirmed these results in
detail\cite{Meden:prb02} and extensions to include finite magnetic
field have been made.\cite{hikihara:cm04} For a single impurity in a
spinless LL, the main result is that for repulsive electron
interactions the impurity ``cuts'' the LL into two semi-infinite
sections, while for attractive electron interactions, the impurity is
irrelevant in the renormalization group (RG) sense and scales to zero
at low energies.  When spin is introduced, the transport can be more
complicated, with the spin and charge sectors possibly behaving
differently. For example, one possibility is that the spin could pass
easily through the impurity while the charge would be reflected.  The
double impurity system exhibits even richer behavior, including
zero-width resonances at zero temperature.  At finite temperature, the
resonance lineshapes exhibit a characteristic non-Lorentzian shape, in
contrast to the case for non-interacting electrons.  When Fermi liquid
leads are attached to the end of a finite length 1-d LL wire, a new
length scale is introduced beyond which the leads play an important
role in the
physics.\cite{Safi:prb95,Maslov:prb95,Lal:prl01,Lal:prb02,Safi:prb99,Meden:prb03,Enss:prb05,Shimizu}
For example, the dc conductance of a clean system is completely
determined by the leads and therefore has a value of $2e^2/h$
independent of the strength of the electron interactions in the wire.

LL theory is based on a picture of interacting electrons in which the
interaction strength is not too great.  In this regime of not too
strong electron interactions, the characteristic exchange energy of
two electrons -- defined by $\hbar$ times the inverse of the time
required for two electrons initially in position eigenstates separated
by the average inter-electron distance to undergo a spin-flip -- is
typically of the same order as the Fermi energy.  However, at low
densities the potential energy grows relative to the kinetic energy
and eventually dominates it for sufficiently low densities when $na_B
\ll 1$, with $n$ the average density of the electrons and the Bohr
radius $a_B=\epsilon \hbar^2/me^2$, with $\epsilon$ the dielectric
constant, $m$ the mass of the electron and $e$ its charge.  At these
low densities, there is a separation of energy scales between the
magnetic exchange energy,\cite{Fogler:prl05,Klironomos:cm05} $J\sim
E_F e^{-\frac{C}{\sqrt{ n a_B}}}$, and the Fermi energy, $E_F=(\pi
\hbar n)^2/8m$. Here $C$ is a positive constant of order unity.  When
the interactions between electrons are very strong, they must tunnel
through one another to exchange, leading to an exponentially small $J$
and a situation where $J \ll E_F$. For $na_B \ll 1$ it is possible to
reach a regime where the temperature $T$ is much larger than the
magnetic exchange energy, but still much less than the Fermi energy:
$J \ll T \ll E_F$.  We refer to this energy scale hierarchy as the
spin incoherent or magnetically incoherent regime.

In this paper, we revisit the transport problem in an interacting 1-d
electron gas, with an eye towards understanding the behavior in the
magnetically incoherent regime.  Recently the magnetically incoherent
regime has been investigated for a clean infinite system by studying
the one particle Green's
function\cite{cheianov03,Cheianov04,Fiete:prl04} and the momentum
distribution function,\cite{Cheianov:cm04} and for finite systems by
studying the influence of incoherent magnetic degrees of freedom on
the momentum resolved tunneling\cite{Fiete:prb05} and on the
conductance of a clean quantum point
contact.\cite{Matveev:prl04,Matveev:prb04} Our main result is that the
transport in an infinite magnetically incoherent electron gas is very
much like that of a spinless LL, except that all the quantum phase
transitions of Kane and Fisher,\cite{Kane:prl92,Kane:prb92} and
Furusaki and Nagaosa \cite{Furusaki:prb93,Nagaosa:prb93}(understood to
be in the limit $J \to 0$, then $T \to 0$) are obtained by replacing
$g$ by $2g_c$, where $g$ is the interaction parameter of the spinless
LL and $g_c$ is the interaction parameter of the charge sector of a LL
theory for electrons with spin.  This result can be understood in the
following physical terms: The condition $T \gg J$ means that the spin
degrees of freedom become non-dynamical in that, within the ``thermal
coherence time'' $t_{th} \sim \hbar/k_B T$, the spin quantum numbers
of individual electrons remain unchanged, since a spin-flip transition
requires a time $t_J \sim \hbar/J \gg t_{th}$ to occur.  Moreover,
because the energy splittings between distinct spin states are
negligible compared to $k_B T$, all spin states are excited with equal
probability.  Hence, dynamically, the electron gas behaves in a
``spinless'' fashion, since the spin degrees of freedom are static and
random, and do not couple to the electron coordinates.  More
specifically, the charge degrees of freedom behave as a LL (because $T
\ll E_F$ ensures only low energy charge excitations are important),
only with effective interaction parameter $2g_c$.  This result can be
established at the level of the Hamiltonian so that the correspondence
$g=2g_c$ in the spin incoherent regime is actually quite general and
applies to any particle-conserving operator. (The single particle
Green's function {\em does not} involve particle conserving operators
and therefore has a qualitatively different form from a spinless
LL.\cite{cheianov03,Cheianov04,Fiete:prl04})

For a finite length wire in the magnetically incoherent regime, the
transport is more subtle.  Matveev has
argued\cite{Matveev:prl04,Matveev:prb04} that for a clean wire the
conductance drops to 1/2 of its zero temperature value giving $e^2/h$
rather than $2e^2/h$. When disorder is present in the spin incoherent
region of the wire, more careful considerations are needed.  We
distinguish between two cases: (i) weak and (ii) strong back scattering
and discuss features of each.

This paper is organized in the following way.  In
Sec.~\ref{sec:quantum_wires} we discuss important physical models,
parameters and limits for quantum wires with low electron density.  In
particular, we show the physics of the incoherent regime is
independent of the {\em range} of electron interactions; the
interactions need only be sufficiently strong to achieve $J \ll E_F$.
In Sec.~\ref{sec:g_identity} we establish the equivalence, summarized
by $g=2g_c$, between a spinless LL and a spin incoherent LL for
particle-conserving charge properties. In Sec.~\ref{sec:leads} we
discuss what the effects of Fermi liquids leads are on the transport
through a finite length quantum wire and then in
Sec.~\ref{sec:crossover} we discuss some details of the crossover from
the magnetically incoherent LL to the familiar spin coherent LL
regime.  Finally, in Sec.~\ref{sec:conclusions} we present our main
conclusions.

\section{Quantum wires at low electron density}
\label{sec:quantum_wires}

It is useful here, in the beginning, to outline the physical
situations where we expect the spin incoherent regime to be realized.
The physics we discuss in this paper is expected to be present
provided the interactions between electrons is sufficiently strong
that there is a separation of magnetic and non-magnetic energy scales
in the problem: $J \ll E_F$. (As we show in
Appendix~\ref{app:LL_finiteT} this regime can also be understood in
terms of a separation of scale in the spin and charge velocities.)  It
is possible to reach this regime with either short range or long range
interactions, so it is not necessary to have a Wigner solid-like
picture in mind. (Strictly speaking, quantum fluctuations destroy the
long range charge order unless the interactions are longer range than
Coulomb.\cite{Schulz:prl93}) However, the Wigner solid picture does
often provide a convenient physical picture for a one dimensional
electron gas, and we will use it to discuss the low
density limit.  In fact, as we discuss below, the effective
Hamiltonian in the Wigner solid limit turns out to be quite universal,
in the sense that its form is independent of the range of the electron
interactions, even down to zero range interactions.

\subsection{Charge and spin Hamiltonians}

With a Wigner solid picture in mind, we can understand the physical
state of the electron gas in classical terms as being dominated by the
Coulomb repulsion between electrons, which forces the electrons to
occupy discrete, evenly spaced positions.  A finite but small kinetic
energy of the electrons implies small displacements from their
equilibrium positions, and the lowest energy displacements are long
wavelength sound modes, or ``phonons".  These displacements can be
described within elasticity theory in terms of the displacement $u(x)$
from equilibrium of the solid at point $x$, and the momentum density,
$p(x)$.\cite{Matveev:prb04} Adding these two contributions to the
energy gives the total energy of the elastic medium
\begin{equation}
\label{eq:H_elas}
H_{\rm elastic}=\int dx \left[\frac{p^2}{2mn}+\frac{1}{2}mns^2(\partial_x u)^2\right]\;,
\end{equation}
where $s=\sqrt{(n/m)(\partial^2E/\partial n^2)}$ is the sound velocity of
 the phonons.\cite{Glazman:prb92}  Here $E$ is the energy of the resting medium per unit length.  

In order to obtain a quantum theory, the Hamiltonian (\ref{eq:H_elas}) can be 
quantized by imposing the commuation relations
$[u(x),p(x')]=i\hbar \delta(x-x')$.  Then new fields can be identified as
\begin{equation}
u(x)=\frac{\sqrt{2}}{n\pi}\theta_c(x),\;\; p(x)=\frac{n\hbar}{\sqrt{2}}\partial_x\phi_c(x)\;,
\end{equation}
which satisfy the commutation relations $[\theta_c(x),\partial_{x'}\phi_c(x')]=i\pi \hbar
\delta(x-x')$.  In terms of these new fields, the Hamiltonian (\ref{eq:H_elas}) becomes
\begin{equation}
\label{eq:H_c}
H_c=\hbar v_c \int \frac{dx}{2\pi}\left[\frac{1}{g_c} (\partial_x\theta_c(x))^2+g_c(\partial_x\phi_c(x))^2\right]\;,
\end{equation}
where 
\begin{equation}
v_c=s,\;\;g_c=\frac{v_F}{s}\;,
\end{equation}
with $v_F=\hbar\pi n/2m$.  The attentive reader will immediately
notice that Eq.~(\ref{eq:H_c}) is just the charge sector Hamiltonian
familiar from LL theory.  Since it is well known that
Eq.~(\ref{eq:H_c}) can be derived for weakly interacting electrons by
linearizing the kinetic energy about the Fermi points, the discussion
above shows that the Hamiltonian (\ref{eq:H_c}) is actually valid for
arbitrary strength interactions. In the rest of this paper, we will
assume that low energy charge states are adequately described by
Eq.~(\ref{eq:H_c}).

So far, we have neglected the spin of the electrons.  Returning to the
Wigner solid picture again, we see that the spin degrees of freedom
will act like a Heisenberg spin chain with lattice spacing equal to
that of the electron spacing.  Virtual hopping of electrons from one
site to another (occupied site) requires the electrons to have
opposite spin, resulting in an effective magnetic exchange for the
spin chain that is antiferromagnetic.  Therefore, the Hamiltonian of
the spin sector behaves as
\begin{equation}
\label{eq:H_s}
H_s= \sum_l J {\vec S_l}\cdot {\vec S_{l+1}} \;,
\end{equation}
where ${\vec S_l}$ is the spin of the $l^{th}$ electron and $J>0$ is the nearest neighbor
exchange energy.  
%In general, $J$ depends on the separation between nearest neighbor
%electrons, and since this distance can fluctuate, there is in general coupling between the
%spin and charge degrees of freedom.  
The spin chain (\ref{eq:H_s}) can be bosonized\cite{Schulz,Giamarchi} 
and the low energy
spin excitations can be computed within a LL theory for the spin
sector.  However, since here we are concerned with the high energy
situation $T\gg J$ (from the point of view of the spin degrees of
freedom), we do not pursue that direction.

\subsection{Long range vs. short range interactions for $T\gg J$}

All quantum wires are ``quasi 1-d'' since there is usually some finite
width to the wire, $w$.  This provides a short range cut off for the
electron interactions at $x\sim w$, so that for $x\lesssim w$,
$V(x)\sim 1/w$.  On the other hand, quantum wires are often gated so
that at electron separations large compared to the distance to the
metallic gate, $d$, (which is always present in experiments) electrons
induce image charges in the gate to produce a dipolar
electron-electron interaction for $x\gtrsim d$, $V(x)\sim d^2/|x|^3$.
A potential consistent with this form is\cite{Hausler:prb02}
\begin{equation}
\label{eq:V_x}
V(x)=\frac{e^2}{\epsilon}\left(\frac{1}{\sqrt{x^2+w^2}}-\frac{1}{\sqrt{x^2+w^2+(2d)^2}}\right)\;,
\end{equation}
where $\epsilon$ is the dielectric constant.  Since we are interested in 
the low density limit where $n^{-1} \gg w$, we can set $w=0$ to obtain an 
approximate form of $V(x)$.  

In the low density limit, we can then argue along the lines of
Ref.~[\onlinecite{Matveev:prb04}].  For $n \ll a_B/d^2$,
Eq.~(\ref{eq:V_x}) shows that the interaction between two particles at
a typical distance of $n^{-1}$ is small compared to their kinetic
energy $\sim E_F$. Here $a_B=\epsilon \hbar^2/me^2$. 
On the other hand, when the distance between
electrons is sufficiently short, $|x| \lesssim n^{-1} (nd^2/a_B)^{1/3}
\ll n^{-1}$, the potential energy dominates the kinetic energy, $V(x)
\gtrsim E_F$.  As a result, at low densities the potential
(\ref{eq:V_x}) can be modeled by the short range potential
\begin{equation}
\label{eq:V_eff}
V^{\rm eff}(x)={\cal V}\delta(x)\;,
\end{equation}
where ${\cal V}$ is chosen to provide the same scattering phase shift as (\ref{eq:V_x}).  

The model (\ref{eq:V_eff}) is equivalent at low energies to the 1-d
Hubbard model. Starting from the Hubbard model, it can be
shown\cite{Ogata:prb90} that in the low density limit with $U/t\to
\infty$ the spin sector takes the form (\ref{eq:H_s}) with the
exchange energy given by\cite{Shiba}
\begin{equation}
\label{eq:J_Hubbard}
J=\frac{4t^2}{U} n_e \left(1-\frac{\sin 2\pi n_e}{2\pi n_e}\right)\;,
\end{equation}
where $n_e$ is the average number of electrons per site. Recall that
we originally motivated the spin Hamiltonian (\ref{eq:H_s}) within the
Wigner solid picture, which relies on sufficiently long range
interactions.  Here, we show that even short range interactions lead
to the spin Hamiltonian (\ref{eq:H_s}).  We can thus view the
Hamiltonian
\begin{equation}
\label{eq:H}
H=H_c+H_s
\end{equation}
as a general form, valid in the energy hierarchy $J \ll E_F$ for
any temperature $T \ll E_F$, including both $J \ll T \ll E_F$
and $T \ll J \ll E_F$.  However, the dependence of $J$ on the
density depends the microscopic details\cite{Fogler:prl05,Klironomos:cm05,Hausler:zpb96}
 of the electron interactions
with the form (\ref{eq:J_Hubbard}) for zero range interactions and
$J\sim E_F e^{-\frac{C}{\sqrt{ n a_B}}}$ for Coulomb interactions.

In the remainder of this paper we will study the implications of the
Hamiltonian (\ref{eq:H}) in the limit $J \ll T \ll E_F$ on the
electrical transport.

\section{Infinitely Long Wire}
\label{sec:g_identity}

In this section we show explicitly that in the limit of strong
interactions and $J \ll T \ll E_F$ the electrons become effectively
spinless (for quantities that do not directly probe spin) and are
governed by a Hamiltonian of the form (\ref{eq:H_c}) with interaction
parameter $g=2g_c$. Here $g$ is the interaction parameter of a
spinless LL and $g_c$ is the interaction parameter of the charge
sector for an electron gas with spin and the same interaction strength
as in the spinless LL. We discuss implications for the transport in
such a magnetically incoherent LL when impurities are present.

\subsection{The relation $g=2g_c$ for $J \ll T \ll E_F$}

We assume that $J \ll T \ll E_F$ and take the limit $J/T\to 0$.  Since
$T \gg J$, the spin degrees of freedom are non-dynamical and the
system behaves as if $J=0$ identically [or equivalently $H_s=0$
identically from (\ref{eq:H_s})].  Thus, the only dynamics is in the
charge sector of the theory. This implies that if we look at any
quantity that does not depend explicitly on spin (conductance or
compressibility, for example) the system behaves as if it were
spinless.  

To show this microscopically, we consider a particular basis of states
for the Hilbert space of the system.  We work in the canonical
ensemble, i.e. with a fixed number of electrons.  Of course, dynamics
in the grand canonical ensemble can be obtained from this by summing
over the sectors with each electron number, since the intrinsic
physical Hamiltonian is anyway number conserving.  For 
fixed electron number,  a convenient real-space basis set is given by
states specifying the position $x_n$ of each electron, and the spin
projection on the $\hat{z}$ axis, $m_n$, in order, from left to right
across the system:
\begin{equation}
  \label{eq:fockbasis}
  |x_1\cdots x_N\rangle |m_1\cdots m_N\rangle =
  c_{m_1}^\dagger(x_1)\cdots c_{m_N}^\dagger(x_N)|0\rangle,
\end{equation}
where $|0\rangle$ is the vacuum state (no particles).

The physics of the spin-incoherent regime is that, within the
thermal coherence time, $t_{th} \sim \hbar/k_B T$, the probability of
a transition between states with different values of $\{ m_n \}$ is
negligible.  Hence, the physics is well-approximated by neglecting
off-diagonal matrix elements in these states.  Moreover, in the same
approximation, for spin-independent interactions, the matrix
elements of $H$ are independent of the $\{ m_n \}$, i.e.
\begin{eqnarray}
  \label{eq:matelts}
&& \langle m'_1\cdots m'_N|\langle x'_1\cdots x'_N|H|x_1 \cdots x_N\rangle
|m_1\cdots m_N\rangle \nonumber \\
&& \approx \langle x'_1\cdots x'_N|H_{sl}|x_1 \cdots
x_N\rangle \delta_{m'_1,m_1}\cdots \delta_{m'_N,m_N},
\end{eqnarray}
where $H_{sl}$ is an effective spinless identical -- and ``hard core''
-- particle Hamiltonian, that governs the (independent) dynamics
within each spin sector.  Note that it is manifestly identical, in
first quantized form, to the original spin-ful Hamiltonian, if the
co\"ordinates of all particles are treated equivalently (as some
spinless particles).

It is important to recognize that this reduction to a spinless
particle problem is extremely general.  In particular, nowhere need we
assume the system is even spatially uniform, only that exchange
processes are everywhere negligible, i.e. $J\ll T$ throughout the
system, that there are no explicit spin-dependent interactions in the
Hamiltonian, and that electrons are not added or removed from the
system during the dynamics.  The equivalence to a spinless problem
continues to hold in the presence of arbitrary potentials, weak links,
etc.

With this understanding, we now address the remaining question of
exactly which spinless theory describes the charge dynamics in the
bulk of the spin incoherent wire.  By the usual LL arguments (as given
above for instance for the Wigner solid), the spinless particle system
is described at low energies ($T,E \ll E_F, E_F r_s$) by the bosonized
Hamiltonian
\bea H_{\rm incoh} &=& \hbar v \int \frac{dx}{2\pi} \left(\frac{1}{g}
  (\partial_x\theta)^2+ g(\partial_x\phi)^2\right)\;, \label{eq:Hinc}
\eea
 with
characteristic ``zero sound'' velocity $v$ and interaction parameter
$g$.  Here we follow one standard convention, in which the
normalization of the fields is fixed by the relation
$\partial_x\theta(x) = \pi \rho(x)$ with $\rho(x)$ the fluctuation in electron density at position $x$, and the commutation relation
$[\theta(x),\partial_{x'}\phi(x')]=i\pi \hbar \delta(x-x')$.  In such
a spinless gas, power-law charge density correlations occur at
wavevectors which are multiples of $Q_{CDW}=2\pi n$ (the reciprocal
lattice vector of the incipient Wigner solid), and we {\sl define}
$Q_{CDW}=2\tilde{k}_F$, which gives $\tilde{k}_F=\pi n$.

Upon crossing over from the spin incoherent regime, $T\gg J$, to the
ultimate low-energy limit, $T\ll J$, we expect the description of the
system to change to the more ``conventional'' spinful LL theory.  This
theory exhibits, as is well-known, spin-charge separation, so the
charge dynamics can be discussed independently.  Again,  standard
arguments give the bosonized effective charge Hamiltonian (including
the effects of rather arbitrary interactions) exhibited in Eq.~(\ref{eq:H_c})
with $v_c= v_F/g_c$, $v_F=\hbar k_F/m$ and $k_F =\pi n/2$.
Normalization is fixed here by $\partial_x\theta_c(x)=\pi
\rho_c(x)/\sqrt{2}$, and $\theta_c,\phi_c$ are taken to obey the same
commutation relations as $\theta,\phi$ above.  The value of the
undetermined ``LL parameter'' $g_c$ depends in detail upon the nature
of interactions in the electron system, but takes the value $g_c=1$
for independent electrons and generally decreases with the strength of
repulsive interactions.

Comparing Eqs.~(\ref{eq:Hinc}) and (\ref{eq:H_c}), one sees a strong
similarity.  In fact, in the limit $J \ll E_F$, we expect they can
indeed be identified.  This is because, with exchange energy so small
compared to the characteristic energies of individual electron (and
hence charge) dynamics, we do not expect the presence of these weak
additional exchange interactions to substantially modify the charge
dynamics itself (except for the emergence of $2k_F$ correlations -- see
Sec.~\ref{sec:crossover} -- which is in fact a ``weak'' effect in this
limit).  Thus, up to corrections of $O(J^2/E_F^2)$, we expect that
{\sl the same} charge Hamiltonian should govern charge dynamics both
for $T\ll J$ and $T\gg J$.  Comparing the different normalization
conventions (relating the fields to the physical charge density
$\rho(x)=\rho_c(x)$), we see that we must equate 
\begin{eqnarray}
  \label{eq:equate}
  \theta & = & \sqrt{2}\theta_c, \\
  \phi & = & \phi_c/\sqrt{2}.
\end{eqnarray}
Requiring identity of $H_{sl}$ and $H_{\rm incoh}$, we immediately
find $v=v_c$ and $g=2g_c$, as promised.  We also obtain $\tilde{k}_F =
2k_F$, which implies the absence of $2k_F=\tilde{k}_F$ oscillating
correlations in the charge density in the spin incoherent regime.
Their emergence at low temperature is discussed in Sec.~\ref{sec:crossover}.

\subsection{Transport through single and double impurities}

Having established in the previous section that for properties that do
not depend explicity on the spin, such as the conductance, a low
density electron gas in the regime $J \ll T \ll E_F$ behaves
effectively as a spinless LL with $g=2g_c$, we are now ready to
address transport properties of an infinitely long wire.  Fortunately,
most of the work has been done for us already by Kane and
Fisher,\cite{Kane:prl92,Kane:prb92} and by Furusaki and
Nagaosa.\cite{Furusaki:prb93,Nagaosa:prb93}  All that needs to be
done is to substitute $g=2g_c$ into their formulas for spinless electrons.
For completeness, we summarize the important results here.

For a single impurity in a spinless LL, the result found earlier was that for
repulsive (attractive) interactions, $g<1$ ($g>1$) the impurity became
relevant (irrelevant) in the renormalization group (RG) sense at the 
lowest energies.  Thus, in the spin
incoherent regime the critical value of  $g_c$ is 1/2.  In particular,
this implies that for $1/2 < g_c <1$ the transparency of a single
barrier should {\em increase rather than decrease} under the RG flows 
for $J \ll T \ll E_F$.  More generally, in the spin incoherent regime,
we expect that for weak tunneling (strong back-scattering)
\begin{equation}
\label{eq:G}
G(T)=\frac{dI}{dV}\propto  T^{(1/g_c)-2}\;,
\end{equation}
and for weak backscattering (strong tunneling)
\begin{equation}
  \label{eq:deltaGinf}
G_0-G(T)\propto T^{2(2g_c-1)}\;,  
\end{equation}
where $G_0=2g_c e^2/h$ is the bare conductance of the spin incoherent ($J=0$) 
wire with no impurities.  

For the two impurity problem the results can also be carried over
directly.  As in the case of the spinless LL with two impurities, the
double impurity nature of the problem will only be relevant at energy
scales below $\sim \hbar v_c/d$ where $d$ is the separation between
the two barriers, since at larger energies the two barriers will not
add coherently.  We remind the reader that the main results from the
spinless LL case are that the double barrier exhibits behavior in
striking contrast to the non-interacting electron liquid, which has
temperature independent resonances with a Lorentzian line shape at low
$T$: For repulsive interactions the LL has non-Lorentzian resonances
with a width that vanishes as $T\to0$. These line shapes are
determined by a universal scaling function:\cite{Kane:prb92} 
\begin{equation}
G(T,\delta)=\tilde G(c\delta/T^\lambda)\;,
\end{equation}
where $\delta$ is a small parameter to tune away from the resonance, $c$ is a dimensionful
constant, and 
\begin{equation}
\lambda=1-2g_c\;.
\end{equation}
Defining $X\equiv c\delta/T^\lambda$, we see that for $X\to0$
\begin{equation}
\tilde G(X)=G_0[1-X^2+{\cal O}(X^4)]\;,
\end{equation}
while for $X\to \infty$,
\begin{equation}
\label{eq:nonlorentz}
\tilde G(X)\approx X^{-1/g_c}\;.
\end{equation}
We note that at low temperature, Eq.~(\ref{eq:nonlorentz}) applies, explicitly showing 
that the tails of the resonance are non-Lorentzian.

\begin{figure}[th]
\includegraphics[width=1.0\linewidth,clip=]{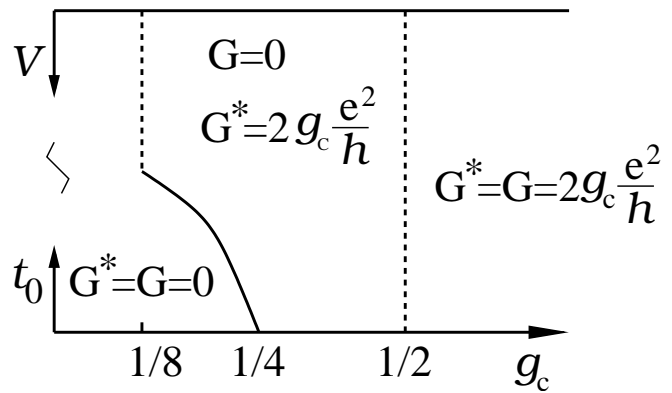}
\caption{\label{fig:phase_diag} Phase diagram 
(in the limit $J\to 0$, then $T\to 0$)
for a spin incoherent Luttinger Liquid with a double barrier structure. 
Here $V$ denotes the strength of the back-scattering for weak impurities,
and $t_0$ the strength of the tunneling for strong impurities.
The conductance off resonance is donoted by $G$ and the conductance on
resonance is denoted by $G^*$.  The dashed line at $g_c=1/2$ is the fixed
line for transmission off resonance separating the regions where electrons
will ($g_c>1/2$) or will not ($g_c <1/2$) propagate through the double barrier structure.  The dashed line at $g_c=1/8$ separates the regions of zero ($g_c<1/8$) and non-zero ($g_c>1/8$) conductance on resonance at vanishing energies.
The solid line between $g_c=1/8$ and 1/4 is a line of Kosterlitz-Thouless
separatricies.  Compare with Fig.~3 of Ref.~[\onlinecite{Kane:prb92}].}
\end{figure}

The most important results are summarized in the phase diagram shown
in Fig.~\ref{fig:phase_diag}.  As in the case of the spinless LL, we
expect there to be a line of Kosterlitz-Thouless separatricies
between the zero conductance and perfect conductance (on resonance)
regions of the phase diagram.

\section{Finite length wire connected to Fermi liquid leads}
\label{sec:leads}

Transport through a finite length segment of a spin incoherent LL
connected to Fermi liquid leads is more subtle than the case discussed
in Sec.~\ref{sec:g_identity} B in which the system is infinite in
length.  Here we have in mind a situation where the finite length
segment is adiabatically connected to the leads so that no
back-scattering occurs due to the slowly changing background potential
itself.  Matveev has recently argued\cite{Matveev:prl04,Matveev:prb04}
that for a clean wire adiabatically connected to FL leads the
conductance is reduced by a factor of 2 when $J \ll T \ll E_F$
compared to $T \ll J, E_F$, resulting in a conductance of $e^2/h$
rather than $2 e^2/h$ for a single mode wire.  The physics of this
result appears to be that when an electron with energy $\sim T$ from
the lead enters the spin incoherent LL portion the spin modes are
reflected because there are no spin states of energy $\sim T \gg J$, while
the charge modes have states up to energy $\sim E_F\gg T$ and are thus able
to pass through.

Here we discuss some additional considerations when there are
impurities in the finite length spin incoherent LL.  We begin with the
simplest case--a single impurity in the center of the wire.  As in the
infinite system, there are two limits to consider: (i) weak
back-scattering and (ii) weak tunneling.  The weak tunneling limit is
the most straightforward of the two.  We also discuss the case of
more than one impurity.

\subsection{Weak tunneling through a finite wire}

Let us first discuss the case of a very strong potential barrier in 
the center of the wire.  In this limit, an infinite one dimensional 
system is cut into two semi-infinite pieces and electrons tunnel 
between these two semi-infinite systems.
The tunneling can be described by the Hamiltonian
\begin{equation}
\label{eq:H_tun1}
H_{tun}=t_0\sum_{\pm}[\Psi^{\dagger}_{1\pm}(0)\Psi_{2\pm}(0)+h.c.],
\end{equation}
where $t_0$ is the tunneling amplitude for an electron to hop from side
1 to side 2.  Using Eq.~(\ref{eq:H_tun1}), the current $I$ through the 
barrier can be computed via Fermi's Golden Rule as
\begin{equation}
I=\frac{et_0^2}{2\pi \hbar}\int d\omega \left[\varrho_1^>(\omega)\varrho_2^<(\omega-eV)-\varrho_1^<(\omega-eV)\varrho_2^>(\omega)\right],
\label{eq:FGR}
\end{equation}
where $\varrho^>$ ($\varrho^<$) is the tunneling density of states for 
adding (removing) an electron {\em at the end of the wire}.  The two 
are related by $\varrho^<(\omega)=\varrho^>(-\omega)$.  The subscripts on 
the density of states refers to the two semi-infinite segments of the 
1-d system.  Clearly then, computing the current depends on knowing 
the tunneling density of states at the end of the wires.  The energy
dependence will depend on the energy itself:  For a wire of length $L$,
for $\hbar \omega \gg \hbar v_c/L$ the Fermi liquid leads will not be
felt and we can use the tunneling density of states of an infinite spin
incoherent wire near a boundary, while for $\hbar \omega \ll \hbar v_c/L$
the energy dependence of the tunneling density of states will be dominated
by the FL leads.

For $eV > \hbar v_c/L$ we can use the tunneling density of states in the 
spin incoherent regime for a semi-infinite system computed in 
Ref.~[\onlinecite{Fiete:prb05}] from the Green's function, 
${\cal G}^>_\sigma(0,\tau)=\langle \Psi_\sigma(0,\tau)\Psi_\sigma^\dagger(0,0)\rangle$, 
\begin{eqnarray}
\label{eq:G0_t}
{\cal G}^>_\sigma(0,\tau)&\sim& \langle e^{i[\phi_c(0,\tau)-\phi_c(0,0)]/\sqrt{2}}\rangle\nonumber \\
&=&  e^{-\langle \tilde \phi_c^2\rangle /4}
 \sim \left(\frac{1}{\tau}\right)^{\frac{1}{2g_c}},
\end{eqnarray}
where $\tilde \phi_c =\phi_c(0,\tau)-\phi_c(0,0)$.  The correlator $\langle \tilde \phi_c^2\rangle$ was
evaluated using the Hamiltonian (\ref{eq:H_c}) subject to the boundary condition $\partial_x\phi_c(x=0)=0$, i.e. that no current passes through the barrier.  (Finite current comes from (\ref{eq:H_tun1}).) See Appendix~\ref{app:corr} for details.

After Fourier transforming the Green's function (\ref{eq:G0_t}) to the frequency domain, the
frequency dependence of the tunneling density of states at the end of the wire is obtained 
\begin{equation}
\label{eq:dos}
\varrho^>(\omega)\sim {\rm Re}[{\cal G}^>(0,\omega)] \sim \omega^{1/(2g_c)-1}\;.
\end{equation}
Note that for $g_c>1/2$ the density of states diverges. This should be
contrasted with the result obtained for the infinite system 
in which $\varrho^>(\omega)\sim \omega^{1/(4g_c)-1}$, and therefore
diverages for $g_c>1/4$.\cite{Fiete:prl04}
Substituting the result (\ref{eq:dos}) into Eq.~(\ref{eq:FGR}) gives the following result for the conductance of the wire (for $T \gtrsim V$, where $V$ is the
voltage)
\begin{equation}
G(T)=\frac{dI}{dV}\propto t_0^2 T^{(1/g_c)-2}\;,\;\;\;\;T,eV \gtrsim
\hbar v_c/L\;. \label{eq:G1}
\end{equation}
This reproduces the expected result of Eq.~(\ref{eq:G}) obtained through
the indentification $g=2g_c$.

Let us now suppose we go to sufficiently low temperatures and voltages
that the ``charge dephasing length'' (this is not strictly proper
terminology, but it will do) is longer than the spin-incoherent wire,
i.e. $T, eV \ll \hbar v_c/L$, but still $J\ll T$.  Then clearly the
charge excitations are modified on these energy scales by the absence of
interactions in the leads.  In an ordinary LL, where the spin and charge
energy scales are comparable, the condition $T \ll \hbar v_c/L$ also
implies $T\ll \hbar v_s/L$, so that in this regime all excitations
become controlled by the leads.  In that more familiar situation, the
tunneling density of states crosses over to a constant value, as
appropriate for a Fermi liquid.  This is not the case for the
spin-incoherent wire.  One may understand this by the fact that, in a
time $\sim \hbar/T$, the spin disturbance created by the tunneling event
has {\sl not} propagated to the leads, and remains within the
spin-incoherent region.

A na\"ive approach to the behavior in this regime is simply to
recalculate ${\cal G}_\sigma^>(0,\tau)$ in Eq.~(\ref{eq:G0_t}) by
assuming a spatially-dependent $g_c(x)$, with $g_c(x) \rightarrow 1$ for
$x$ outside the wire (in the leads).  This gives a decay at long times,
${\cal G}_\sigma^>(0,\tau)\sim 1/\sqrt{\tau}$.  Matching this to
Eq.~(\ref{eq:G1}) for $T\sim \hbar v_c/L$, one finds
\begin{equation}
G=\frac{dI}{dV}\propto t_0^2 \left(\frac{\hbar v_c}{L}\right)^{1/g_c-1}
\frac{1}{T} \;,\;\;\;\;T,eV \lesssim \hbar v_c/L\;,
\end{equation}
valid so long as one is in the week tunneling limit, $G \ll e^2/h$.
Remarkably, the conductance does not become constant for such
temperatures, but actually {\sl diverges more strongly} with decreasing
temperature than for an infinite spin-incoherent wire.  This is because,
since the spin excitation created by the tunneling event is still within
the wire, the enhancement of the density of states is still operative,
while the competing suppression of the density of states due to
``charging'' is no longer in effect once the charge disturbance has
exited the wire.  A crossover to true Fermi-liquid behavior is therefore
only expected once the spin disturbance has had time to reach the leads,
requiring $T\ll J$.

Similar reasoning applies to the case when two impurities are present
in the short wire, but it is more involved.  We expect the same
results as we found for the infinite spin incoherent wire for $T,eV
\gtrsim \hbar v_c/L$, that is the Kane-Fisher and Furusaki-Nagaosa
results with $g=2g_c$.  For low energies, $T,eV \lesssim \hbar v_c/L$,
the leads again play an important role in the conductance, the precise
nature of which remains to be determined.

\subsection{Weak back-scattering in a finite wire}

For $J \ll T \ll E_F$, and a weak impurity, we are asking about
corrections to the simpler problem, attacked by Matveev, for the clean
wire.  It is clear that, even in the clean case, the spin-incoherent
region constitutes a strong deviation from the usual regime of ideal
conduction quantization, $G_{\rm ideal}=2e^2/h$.  Matveev has given
arguments that suggest the conductance is reduced to
\begin{equation}
  \label{eq:mvscale}
G(T)= \frac{e^2}{h}+\frac{e^2}{h}F(J/T)\;,  
\end{equation}
where $F(x)\to 0$ for $x\to 0$ and  $F(x)\to 1$ for $x\to \infty$, so
that $G(T) \approx e^2/h$ for $J \ll T$ deep in the spin incoherent
limit.  

Strictly speaking, the impurity corrections in this limit should be
calculated using perturbation theory starting from known correlation
functions of the clean ``Matveev problem''.  Unfortunately, it is not
clear from the analysis of
Refs.~[\onlinecite{Matveev:prl04},\onlinecite{Matveev:prb04}] how to
carry out such a perturbative calculation.  Lacking this, we do not
discuss this problem in detail.  A general remark is that, in this
limit, the corrections to the conductance due to the impurity are {\sl
small}, and hence require precision to observe.  Once the corrections
are no longer small (which will occur at low enough temperature, for
$g_c<1/2$, provided the thermal length does not first exceed the wire
length), the perturbative approach has broken down.

Can we guess the nature of the first perturbative correction?  Let us
assume that the (charge) thermal length is less than the length of the
(spin incoherent) wire, $\hbar v_c/T<L$, for simplicity.  A na\"ive
extension of the arguments of
Refs.~[\onlinecite{Matveev:prl04},\onlinecite{Matveev:prb04}] then
gives a suggestion.  The arguments therein proceed by determining the
power radiated to infinity by the spin-charge separated modes in a
bosonized formulation of the leads.  This dissipated power, calculated
for an imposed (charge) current $I$, is proportional, by definition,
to $I^2 R$, where $R$ is the physical resistance.  In this
formulation, it appears that the charge and spin resistances add,
$R=R_c + R_s$, since the charge and spin sectors provide separate
channels for energy to radiate, and we are interested in the total
rate of radiation.  Na\"ively, a weak impurity in the spin incoherent
region, which has the form of a potential, couples only to the charge
density.  We may therefore na\"ively estimate its effect by
calculating the leading order increase in $R_c$ by the impurity.  Thus,
one expects a (small) additive contribution to the resistance, whose
scaling form (e.g. power-law dependence on temperature etc.) is that
of a a spinless LL with spinless Luttinger parameter $g=2g_c$, i.e the
right-hand-side of Eq.~(\ref{eq:deltaGinf}).

\section{Crossover to spin coherent regime}
\label{sec:crossover}

In order to better understand the physics (including and beyond
electrical transport) to be expected in the spin incoherent LL regime,
it is worthwhile to elaborate on some features of the crossover
between $J \ll T \ll E_F$ and $T \ll J \ll E_F$.  A discussion of the
approach to the spin incoherent LL from a finite temperature LL is
discussed in Appendix~\ref{app:LL_finiteT}.  A discussion of the
changes to the spectral function (obtained by Fourier transforming the
Green's function, for example) are given in
Refs.[\onlinecite{Fiete:prl04,Fiete:prb05}].  The main result is that
the propagating spin mode of the LL theory is lost and the charge mode
excitation is broadened in momentum space to an amount of order the
Fermi wavevector. For short-range interactions there is also a shift
in the ``edge'' of the momentum distribution from $k_F=\pi n/2$ to
$\tilde k_F=2k_F=\pi n$ as $T$ goes from below to 
above $J$.\cite{Cheianov:cm04}
Moreover, there is also a crossover from $2k_F$ to $4k_F$ oscillations
in the density-density correlation function.\cite{Fiete:unpublished}
This can be seen by introducing magneto-elastic coupling in the spin
Hamiltonian (\ref{eq:H_s}) by making the magnetic exchange $J$ depend
on the displacement from equilibrium of neighboring electrons (assuming
the Wigner solid picture we discussed earlier), 
$u_{l+1}-u_l$, as
\begin{equation}
J_l=J_0 +J_1(u_{l+1}-u_l) + {\cal O}((u_{l+1}-u_l) ^2)\;.
\end{equation}
To second order in $J_1$ one finds
\bea
\label{eq:rho_rho_2}
 \langle \rho(x)\rho(x')\rangle^{(2)} \propto (J_1)^2\sum_{l,l'}\int_0^{\beta} d\tau_1 \int_0^{\tau_1}d\tau_2 \nonumber \\ \langle {\vec S_{l+1}}\cdot{\vec S_l}(\tau_1) {\vec S_{l'+1}}\cdot{\vec S_{l'}}(\tau_2)\rangle^{(0)} \langle \rho(x)\rho(x')\rangle^{(0)}\;,\nonumber \\
\eea
where $\langle \rho(x)\rho(x')\rangle^{(0)}$ is the density-density 
correlation function for $J=J_0$ identically and 
$\langle {\vec S_{l+1}}\cdot{\vec S_l}(\tau_1) {\vec S_{l'+1}}\cdot{\vec S_{l'}}(\tau_2)\rangle^{(0)}$ is the dimer correlation function for $J=J_0$ 
identically and $\rho(x)=\sum_l\delta(x-al-u_l)$. In the limit of strong interactions ($J \ll E_F$) 
considered here, 
$\langle \rho(x)\rho(x')\rangle^{(0)}$ contains only $4k_F$ oscillations.
However, at second order the dimer correlation function enters and for
$T \ll J$ a small lattice distortion can lower the energy by allowing
singlet pairs to form on adjacent pairs of sites.  This produces a $2k_F$
oscillation in the dimer correlation function which enters the density-density correlation function at second order and thus produces $2k_F$ oscillations in that quantity as well. When $T \gg J$ the dimer correlations are lost
and only the $4k_F$ oscillations will remain in the density-density 
correlation function.

The loss of $2k_F$ oscillations in the density-density correlation
function as $T$ rises above $J$ will have implications for electrical
transport and drag between parallel quantum wires.  In particular, if
there is some potential $V(x)$ acting on the electrons in the wire
(either static due to impurities, or dynamic due to electrons in
another nearby wire in the drag geometry) with significant $2k_F$
Fourier components coupling to the electron density, these components
of the density oscillations will be lost when $T \gg J$.  This could
lead to a sharp temperature dependence of the electrical transport.

\section{Conclusions}
\label{sec:conclusions}

The main conclusion of this paper is that the physics of
particle-conserving quantities that do not explicity depend on spin
are described by a spinless Hamiltonian with $g=2g_c$, in the regime
$J \ll T \ll E_F$ where $g$ is the interaction parameter of the
spinless LL and $g_c$ is the interaction parameter of the charge
sector in the usual LL theory of electrons with spin that describes
when $T\ll J$. Physically this follows from the condition $J \ll T$,
which renders the spins effectively non-dynamical.  The condition $T
\ll E_F$ allows the charge sector to be described by an effective low
energy LL theory.

As an application, we discuss single and double impurity problems of
an infinitly long 1-d electron gas and a finite length system coupled
to Fermi liquid leads. In the infinite case, all of the phase diagrams
for a spinless LL can be directly used to obtain the behavior of the
spin incoherent 1-d system by using the identification $g=2g_c$.  For
a finite length spin incoherent LL of length $L$, and for energies
larger than $\hbar v_c/L$ the transport behaves much like the infinite
case.  However, for lower energies the effects of the FL leads
dominate the transport.

We have discussed how the condition $J \ll E_F$ implies that the
electron interactions must be very strong, and that this separation of
magnetic and non-magnetic energy scales does not depend on the range
of the interactions.  We have also discussed how a 1-d Wigner solid picture
of electrons at low density provides a clear physical picture of how
the physics we discuss in this paper may arise, and we have shown how
the physical results of interest obtained within the Wigner solid picture
are actually quite general since they can also be shown to hold for models with very strong, but short range interactions, such as the Hubbard model.

We have also discussed general features of the spin incoherent regime
and which properties are expected to change as a function of
temperature when $J\ll T$ or when $J \gg T$.  For example, the $2k_F$
oscillations of the density-density correlation function are lost at
$T \gg J$ and this may affect the coupling of the density to a
background potential and hence the transport or any other quantity
that depends on density variations in the electron gas.

\acknowledgments

We thank M. P. A. Fisher, B. I. Halperin, W. Hofstetter, A. W. W.
Ludwig, K. Matveev, and D. Polyakov for discussions and A. Furusaki
for helpful comments on the manuscript. K.L.H. would like to thank K. Matveev
for insightful comments on the spin incoherent regime. G.A.F. and
K.L.H. were supported by NSF PHY99-07949, L.B. and G.A.F. by NSF
DMR-9985255 and the Packard Foundation, and K.L.H.  by CIAR, FQRNT,
and NSERC.  K.L.H. thanks the KITP at UC Santa Barbara for hospitality
where part of this work was completed.

\appendix
\section{Finite temperature Luttinger Liquid theory for $v_s \ll v_c$}
\label{app:LL_finiteT}

The approach to the spin incoherent regime $J \ll T \ll E_F$ from the
LL state can be understood in the limit of vanishing spin velocity
\cite{spin_v_comment}
relative to charge velocity, $v_s/v_c \to 0$ at finite
temperature. 

We assume that the spin and charge Hamiltonians are given by the LL forms 
\begin{equation}
\label{eq:H_c_app} 
H_c=\hbar v_c \int
\frac{dx}{2\pi}\left[\frac{1}{g_c}
(\partial_x\theta_c(x))^2+g_c(\partial_x\phi_c(x))^2\right]\;,
\end{equation} 
\begin{equation} \label{eq:H_s_app} H_s=\hbar v_s \int
\frac{dx}{2\pi}\left[\frac{1}{g_s}
(\partial_x\theta_s(x))^2+g_s(\partial_x\phi_s(x))^2\right]\;,
\end{equation} 
where the charge(c) and spin(s) fields are
\begin{eqnarray}
\theta_c=\frac{1}{\sqrt{2}}\left(\theta_\uparrow+\theta_\downarrow\right),\;
\theta_s=\frac{1}{\sqrt{2}}\left(\theta_\uparrow-\theta_\downarrow\right)\;,\\
\phi_c=\frac{1}{\sqrt{2}}\left(\phi_\uparrow+\phi_\downarrow\right),\;
\phi_s=\frac{1}{\sqrt{2}}\left(\phi_\uparrow-\phi_\downarrow\right)\;,
\end{eqnarray}
so that $[\partial_x\theta_\alpha(x),\phi_{\beta}(x')]=-i\pi
\delta(x-x')\delta_{\alpha,\beta}$, where $\alpha,\beta=c$ or $s$.

Let us consider the one particle Green's function in
imaginary time: 
\begin{equation} {\cal
G}_\sigma(x,\tau)=\langle\Psi_\sigma(x,\tau)\Psi_\sigma^\dagger(0,0)\rangle\;.
\end{equation} 
where the average is taken at finite
temperature. Neglecting the rapidly oscillating pieces coming from
$R\to L$ and $L\to R$ scattering, we have 
\begin{equation} 
{\cal
G}_\sigma(x,\tau)=\langle\Psi^R_\sigma(x,\tau)\Psi_\sigma^{R\dagger}(0,0)\rangle+
\langle\Psi^L_\sigma(x,\tau)\Psi_\sigma^{L\dagger}(0,0)\rangle\;.
\end{equation} 
Substituting 
\begin{eqnarray}
\label{eq:Psi_R}
\Psi_\pm^R(x)=\frac{1}{\sqrt{2\pi a}} e^{i (k_F x+\theta_c(x)/\sqrt{2})} e^{i \phi_c(x)/\sqrt{2}}\nonumber \\
\times e^{\pm i (\theta_s(x)/\sqrt{2})} e^{\pm i \phi_s(x)/\sqrt{2}}\;,
\end{eqnarray}
\begin{eqnarray}
\label{eq:Psi_L}
\Psi_\pm^L(x)=\frac{1}{\sqrt{2\pi a}} e^{-i  (k_F x+\theta_c(x)/\sqrt{2})} e^{i \phi_c(x)/\sqrt{2}}\nonumber \\
\times e^{\mp i  (\theta_s(x)/\sqrt{2})} e^{\pm i \phi_s(x)/\sqrt{2}}\;,
\end{eqnarray}
where $+$ ($-$) refers to spin $\uparrow$ ($\downarrow$),
using the imaginary time path integral
representation, and the Gaussian action that follows from
(\ref{eq:H_c_app}) and (\ref{eq:H_s_app}), we find 
\begin{eqnarray}
\label{eq:G_corr}
\langle\Psi^R_\pm(x,\tau)\Psi_\pm^{R\dagger}(0,0)\rangle\sim e^{ik_Fx}
e^{-\frac{1}{4}\langle(\tilde \theta_c(x,\tau)^2\rangle}
e^{-\frac{1}{4}\langle(\tilde \phi_c(x,\tau)^2\rangle}\nonumber \\
e^{-\frac{1}{2}\langle \tilde \theta_c (x,\tau) \tilde
\phi_c(x,\tau)\rangle} \nonumber \\ \times
e^{-\frac{1}{4}\langle(\tilde \theta_s(x,\tau)^2\rangle}
e^{-\frac{1}{4}\langle(\tilde \phi_s(x,\tau)^2\rangle}\nonumber \\
e^{\mp \frac{1}{2}\langle \tilde \theta_s (x,\tau) \tilde
\phi_s(x,\tau)\rangle}\nonumber \\ 
\end{eqnarray} 
where the $\tilde
\theta_c(x,\tau)=\theta_c(x,\tau)-\theta_c(0)$, etc.  The
corresponding formula for
$\langle\Psi^L_\pm(x,\tau)\Psi_\pm^{L\dagger}(0,0)\rangle$ has
$k_F\to-k_F$ and $\langle \tilde \theta(x,\tau) \tilde
\phi(x,\tau)\rangle \to - \langle \tilde \theta (x,\tau) \tilde
\phi(x,\tau)\rangle$ for both the spin and charge sectors.  All of the
correlators in (\ref{eq:G_corr}) can be evaluated by doing Gaussian
integrals.  They are 
\begin{eqnarray} 
\label{eq:theta_corr} 
\langle
\tilde \theta_c(x)^2\rangle=2 \pi \hbar g_c \int \frac{dk}{2\pi}
T\sum_{\omega_n}\frac{1-e^{i(kx-\omega_n\tau)}}{\omega_n^2+v_c^2k^2}\nonumber
\\ =\frac{g_c}{2}\ln\left[\frac{\cosh(2\pi Tx/v_c)-\cos(2\pi
T\tau)}{(2\pi T/v_c\Lambda)^2}\right]\;.  
\end{eqnarray} 
where
$\Lambda$ is a large momentum cut off.  The other correlators are
computed likewise 
\begin{equation} 
\label{eq:phi_corr} 
\langle \tilde \phi_c(x)^2\rangle=\frac{1}{2g_c}\ln\left[\frac{\cosh(2\pi
Tx/v_c)-\cos(2\pi T\tau)}{(2\pi T/v_c\Lambda)^2}\right]\;,
\end{equation} 
and 
\begin{equation} 
\label{eq:thetaphi_corr} 
\langle\tilde \theta_c(x)\tilde \phi_c(0)\rangle=
\frac{1}{2}\ln\left[\frac{\tanh(\pi T x/v_c)+i
\tan(\pi T \tau)}{\tanh(\pi T x/v_c)-i \tan(\pi T \tau)}\right]\;,
\end{equation} 
with the corresponding formulas for the spin sector
obtained by replacing $v_c\to v_s$ and $g_c\to g_s$.

As the density of the electron gas is lowered, the ratio $v_s/v_c\to 0$
with decreasing density for $n^{-1}=a>>a_B$. Consider the equal time Green's function where $\tau=0$.  According to Eqs.~(\ref{eq:theta_corr})-(\ref{eq:thetaphi_corr}),
\begin{eqnarray}
{\cal G}_\sigma(x,0)\sim (e^{ik_F x} +e^{-ik_F x}) e^{-\frac{1}{4}\langle(\tilde \theta_c(x)^2\rangle} e^{-\frac{1}{4}\langle(\tilde \phi_c(x)^2\rangle} \nonumber \\
\times e^{-\frac{1}{4}\langle(\tilde \theta_s(x)^2\rangle} e^{-\frac{1}{4}\langle(\tilde \phi_s(x)^2\rangle}\nonumber \\
=(e^{ik_F x} +e^{-ik_F x}) \left(\frac{(2\pi T/v_c\Lambda)^2}{\cosh(2\pi Tx/v_c)-1}\right)^{\frac{g_c+g_c^{-1}}{8}}\nonumber \\
\times \left(\frac{(2\pi T/v_s\Lambda)^2}{\cosh(2\pi Tx/v_s)-1} \right)^{\frac{g_s+g_s^{-1}}{8}}\;.
\end{eqnarray}
For very small temperatures we recover the LL result,
\begin{eqnarray}
{\cal G}_\sigma(x,0)\sim (e^{ik_F x} +e^{-ik_F x}) \left(\frac{(1/\Lambda)^2}{x^2}\right)^{\frac{g_c+g_c^{-1}}{8}}\nonumber \\ \times \left(\frac{(1/\Lambda)^2}{x^2}\right)^{\frac{g_s+g_s^{-1}}{8}}\;.
\end{eqnarray}
For finite temperatures, the Green's function is cut off at large $x$ when $2\pi Tx/v\gg1$ where $v$ is either the charge or spin velocity.  Since for low electron density, we have $v_s\ll v_c$, it is possible to have $2\pi Tx/v_s\gg1$, but $2\pi Tx/v_c\ll1$. In this parameter range, the Green's function behaves
as 
\begin{equation}
{\cal G}_\sigma(x,0)\sim \left(\frac{(1/\Lambda)^2}{x^2}\right)^{\frac{g_c+g_c^{-1}}{8}} \left(\frac{2\pi T}{v_s\Lambda}\right)^{\frac{g_s+g_s^{-1}}{4}}e^{-|x|/\xi_s}\;,
\end{equation}
where the spin coherence length is given by $\xi_s=4 v_s/(\pi (g_s+g_s^{-1}) T)$.  Luttinger liquid theory is valid only on length scales long compared to the lattice spacing, which in the present case is the mean electron separation.  Thus, we expect that the minimum coherence length occurs when $\xi_s^* \approx a$.  This implies a $T^* \sim v_s/a \sim J$ where $J$ is the characteristic interparticle exchange energy.  LL liquid theory therefore only applies down to temperatures in the range $J\sim T \ll E_F$.  When $J \ll T$ and $T\ll E_F$ we expect qualitatively new physics.  It is precisely this regime
that we focus on in this paper. 

\section{Evaluatation of correlators appearing in bosonized formulas for infinite and semi-infinite systems}
\label{app:corr}

For an infinite system, the $\theta_c$ and $\phi_c$ correlators are readily evaluated using the
Hamiltonian (\ref{eq:H_c_app}) and then going to a path integral representation which results
in Gaussian integrals.  The finite temperature correlators have already been computed in 
Eqs.~(\ref{eq:theta_corr})-(\ref{eq:thetaphi_corr}), from which the zero temperature results
(valid for $T\ll E_F$) are readily extracted
\begin{eqnarray}
\langle \tilde \theta_c(x)^2\rangle&=& \frac{g_c}{2} \ln\left[(x^2+(v_c\tau)^2)/a^2\right]\;, \\
\label{eq:corr_phi_inf}
\langle \tilde \phi_c(x)^2\rangle&=& \frac{1}{2g_c} \ln\left[(x^2+(v_c\tau)^2)/a^2\right]\;,\\
\langle\tilde \theta_c(x)\tilde \phi_c(0)\rangle&=&
\frac{1}{2}\ln\left[\frac{x+i  v_c \tau}{x-i v_c \tau}\right]\;.
\end{eqnarray} 

An important correlator that appears in the problem of strong barriers is $\langle \tilde \phi_c(x)^2\rangle$ at the end of a semi-finite wire. This correlator can be readily evaluated
using the following expansion of the fields in a Fourier series:

\begin{eqnarray}
\label{eq:theta_FT}
\theta_c(x,\tau)=\sum_{m=1}^\infty i \sqrt{\frac{g_c}{m}}\sin\left(\frac{m\pi x}{L}\right)(b_m e^{-\omega_m \tau}-b^\dagger_m e^{\omega_m \tau})\nonumber \\
+\theta^{(0)}(x)\;\;\;\;\;
\end{eqnarray}
\begin{eqnarray}
\label{eq:phi_FT}
\phi_c(x,\tau)=\sum_{m=1}^\infty\sqrt{\frac{1}{g_cm}}\cos\left(\frac{m\pi x}{L}\right)(b_m e^{-\omega_m \tau}+b^\dagger_m e^{\omega_m \tau})\nonumber \\
+\Phi_c\;,\;\;\;\;\;\;\;\;\;\;
\end{eqnarray}
where the zero mode term $\theta_c^{(0)}(x)\equiv
\frac{x}{L}\frac{\pi}{\sqrt{2}} N$, the $b_m$ satisfy
$[b_m,b^\dagger_{m'}]=\delta_{mm'}$ and the operators $N$ and $\Phi_c$
satisfy $[N,\Phi_c]=1$. We have assumed that the electrons are
confined to a wire of length $L$.  The fields must satisfy the
boundary conditions $\partial_x\phi_c(x=0,L)=0$ and
$\theta_c(L)-\theta_c(0)=\pi N/\sqrt{2}$ where $N$ is the total number
of electrons.

Using Eq.~(\ref{eq:phi_FT}) it is readily found that at the boundary
\begin{equation}
\label{eq:corr_phi_bd}
\langle \tilde \phi_c(x=0)^2\rangle^{\rm boundary}= \frac{2}{g_c} \ln\left[v_c\tau/a\right]\;.
\end{equation}

By comparing the boundary value of the correlator
(\ref{eq:corr_phi_bd}) with the expression for the infinite system for
$x=0$ (\ref{eq:corr_phi_inf}), we see that the effect of the boundary
is to ``double'' the exponent of $e^{-\langle \tilde
\phi_c(x=0)^2\rangle}$.  This exponent ``doubling'' can be understood
quite simply by going to a chiral fermion basis valid on the whole
real axis, rather than the basis we have used here so far, which is
only valid for the positive half-line.  The resulting chiral action
contains an overall factor of 1/2 difference from the non-chiral case
and this factor translates into the factor of 2 that ``doubles'' the
exponent of the infinite case.

%\bibliography{fiete_nanowires.bib}

\begin{thebibliography}{39}
\expandafter\ifx\csname natexlab\endcsname\relax\def\natexlab#1{#1}\fi
\expandafter\ifx\csname bibnamefont\endcsname\relax
  \def\bibnamefont#1{#1}\fi
\expandafter\ifx\csname bibfnamefont\endcsname\relax
  \def\bibfnamefont#1{#1}\fi
\expandafter\ifx\csname citenamefont\endcsname\relax
  \def\citenamefont#1{#1}\fi
\expandafter\ifx\csname url\endcsname\relax
  \def\url#1{\texttt{#1}}\fi
\expandafter\ifx\csname urlprefix\endcsname\relax\def\urlprefix{URL }\fi
\providecommand{\bibinfo}[2]{#2}
\providecommand{\eprint}[2][]{\url{#2}}

\bibitem[{\citenamefont{Haldane}(1981)}]{haldane81}
\bibinfo{author}{\bibfnamefont{F.}~\bibnamefont{Haldane}}, \bibinfo{journal}{J.
  Phys. C} \textbf{\bibinfo{volume}{14}}, \bibinfo{pages}{2585}
  (\bibinfo{year}{1981}).

\bibitem[{\citenamefont{Voit}(1995)}]{Voit:rpp95}
\bibinfo{author}{\bibfnamefont{J.}~\bibnamefont{Voit}}, \bibinfo{journal}{Rep.
  Prog. Phys.} \textbf{\bibinfo{volume}{58}}, \bibinfo{pages}{977}
  (\bibinfo{year}{1995}).

\bibitem[{\citenamefont{Auslaender et~al.}(2005)\citenamefont{Auslaender,
  Steinberg, Yacoby, Tserkovnyak, Halperin, Baldwin, Pfeiffer, and
  West}}]{Steinberg:sci05}
\bibinfo{author}{\bibfnamefont{O.~M.} \bibnamefont{Auslaender}},
  \bibinfo{author}{\bibfnamefont{H.}~\bibnamefont{Steinberg}},
  \bibinfo{author}{\bibfnamefont{A.}~\bibnamefont{Yacoby}},
  \bibinfo{author}{\bibfnamefont{Y.}~\bibnamefont{Tserkovnyak}},
  \bibinfo{author}{\bibfnamefont{B.~I.} \bibnamefont{Halperin}},
  \bibinfo{author}{\bibfnamefont{K.~W.} \bibnamefont{Baldwin}},
  \bibinfo{author}{\bibfnamefont{L.~N.} \bibnamefont{Pfeiffer}},
  \bibnamefont{and} \bibinfo{author}{\bibfnamefont{K.~W.} \bibnamefont{West}},
  \bibinfo{journal}{Science} \textbf{\bibinfo{volume}{308}},
  \bibinfo{pages}{88} (\bibinfo{year}{2005}).

\bibitem[{\citenamefont{Ishii et~al.}(2003)\citenamefont{Ishii, Kataura,
  Shiozawa, Yoshioka, Otsubo, Takayama, Miyahara, Suzuki, Achiba, Nakatake
  et~al.}}]{ishii:nat03}
\bibinfo{author}{\bibfnamefont{H.}~\bibnamefont{Ishii}},
  \bibinfo{author}{\bibfnamefont{H.}~\bibnamefont{Kataura}},
  \bibinfo{author}{\bibfnamefont{H.}~\bibnamefont{Shiozawa}},
  \bibinfo{author}{\bibfnamefont{H.}~\bibnamefont{Yoshioka}},
  \bibinfo{author}{\bibfnamefont{H.}~\bibnamefont{Otsubo}},
  \bibinfo{author}{\bibfnamefont{Y.}~\bibnamefont{Takayama}},
  \bibinfo{author}{\bibfnamefont{T.}~\bibnamefont{Miyahara}},
  \bibinfo{author}{\bibfnamefont{S.}~\bibnamefont{Suzuki}},
  \bibinfo{author}{\bibfnamefont{Y.}~\bibnamefont{Achiba}},
  \bibinfo{author}{\bibfnamefont{M.}~\bibnamefont{Nakatake}},
  \bibnamefont{et~al.}, \bibinfo{journal}{Nature}
  \textbf{\bibinfo{volume}{426}}, \bibinfo{pages}{540} (\bibinfo{year}{2003}).

\bibitem[{\citenamefont{Bockrath et~al.}(1999)\citenamefont{Bockrath, Cobden,
  Lu, Rinzler, Smalley, Balents, and McEuen}}]{Bockrath:nat99}
\bibinfo{author}{\bibfnamefont{M.}~\bibnamefont{Bockrath}},
  \bibinfo{author}{\bibfnamefont{D.~H.} \bibnamefont{Cobden}},
  \bibinfo{author}{\bibfnamefont{J.}~\bibnamefont{Lu}},
  \bibinfo{author}{\bibfnamefont{A.~G.} \bibnamefont{Rinzler}},
  \bibinfo{author}{\bibfnamefont{R.~E.} \bibnamefont{Smalley}},
  \bibinfo{author}{\bibfnamefont{L.}~\bibnamefont{Balents}}, \bibnamefont{and}
  \bibinfo{author}{\bibfnamefont{P.~L.} \bibnamefont{McEuen}},
  \bibinfo{journal}{Nature} \textbf{\bibinfo{volume}{397}},
  \bibinfo{pages}{598} (\bibinfo{year}{1999}).

\bibitem[{\citenamefont{Yao et~al.}(1999)\citenamefont{Yao, Postma, Balents,
  and Dekker}}]{Yao:nat99}
\bibinfo{author}{\bibfnamefont{Z.}~\bibnamefont{Yao}},
  \bibinfo{author}{\bibfnamefont{H.~W.~C.} \bibnamefont{Postma}},
  \bibinfo{author}{\bibfnamefont{L.}~\bibnamefont{Balents}}, \bibnamefont{and}
  \bibinfo{author}{\bibfnamefont{C.}~\bibnamefont{Dekker}},
  \bibinfo{journal}{Nature} \textbf{\bibinfo{volume}{402}},
  \bibinfo{pages}{273} (\bibinfo{year}{1999}).

\bibitem[{\citenamefont{Kane and Fisher}(1992{\natexlab{a}})}]{Kane:prl92}
\bibinfo{author}{\bibfnamefont{C.~L.} \bibnamefont{Kane}} \bibnamefont{and}
  \bibinfo{author}{\bibfnamefont{M.~P.~A.} \bibnamefont{Fisher}},
  \bibinfo{journal}{Phys. Rev. Lett.} \textbf{\bibinfo{volume}{68}},
  \bibinfo{pages}{1220} (\bibinfo{year}{1992}{\natexlab{a}}).

\bibitem[{\citenamefont{Kane and Fisher}(1992{\natexlab{b}})}]{Kane:prb92}
\bibinfo{author}{\bibfnamefont{C.~L.} \bibnamefont{Kane}} \bibnamefont{and}
  \bibinfo{author}{\bibfnamefont{M.~P.~A.} \bibnamefont{Fisher}},
  \bibinfo{journal}{Phys. Rev. B} \textbf{\bibinfo{volume}{46}},
  \bibinfo{pages}{15233} (\bibinfo{year}{1992}{\natexlab{b}}).

\bibitem[{\citenamefont{Furusaki and
  Nagaosa}(1993{\natexlab{a}})}]{Furusaki:prb93}
\bibinfo{author}{\bibfnamefont{A.}~\bibnamefont{Furusaki}} \bibnamefont{and}
  \bibinfo{author}{\bibfnamefont{N.}~\bibnamefont{Nagaosa}},
  \bibinfo{journal}{Phys. Rev. B} \textbf{\bibinfo{volume}{47}},
  \bibinfo{pages}{4631} (\bibinfo{year}{1993}{\natexlab{a}}).

\bibitem[{\citenamefont{Furusaki and
  Nagaosa}(1993{\natexlab{b}})}]{Nagaosa:prb93}
\bibinfo{author}{\bibfnamefont{A.}~\bibnamefont{Furusaki}} \bibnamefont{and}
  \bibinfo{author}{\bibfnamefont{N.}~\bibnamefont{Nagaosa}},
  \bibinfo{journal}{Phys. Rev. B} \textbf{\bibinfo{volume}{47}},
  \bibinfo{pages}{3827} (\bibinfo{year}{1993}{\natexlab{b}}).

\bibitem[{\citenamefont{Meden et~al.}(2002)\citenamefont{Meden, Metzner,
  Schollw\"ok, and Sch\"onhammer}}]{Meden:prb02}
\bibinfo{author}{\bibfnamefont{V.}~\bibnamefont{Meden}},
  \bibinfo{author}{\bibfnamefont{W.}~\bibnamefont{Metzner}},
  \bibinfo{author}{\bibfnamefont{U.}~\bibnamefont{Schollw\"ok}},
  \bibnamefont{and}
  \bibinfo{author}{\bibfnamefont{K.}~\bibnamefont{Sch\"onhammer}},
  \bibinfo{journal}{Phys. Rev. B} \textbf{\bibinfo{volume}{65}},
  \bibinfo{pages}{045318} (\bibinfo{year}{2002}).

\bibitem[{\citenamefont{Hikihara et~al.}(2005)\citenamefont{Hikihara, Furusaki,
  and Matveev}}]{hikihara:cm04}
\bibinfo{author}{\bibfnamefont{T.}~\bibnamefont{Hikihara}},
  \bibinfo{author}{\bibfnamefont{A.}~\bibnamefont{Furusaki}}, \bibnamefont{and}
  \bibinfo{author}{\bibfnamefont{K.~A.} \bibnamefont{Matveev}},
  \bibinfo{journal}{Phys. Rev. B} \textbf{\bibinfo{volume}{72}},
  \bibinfo{pages}{035301} (\bibinfo{year}{2005}).

\bibitem[{\citenamefont{Safi and Schulz}(1995)}]{Safi:prb95}
\bibinfo{author}{\bibfnamefont{I.}~\bibnamefont{Safi}} \bibnamefont{and}
  \bibinfo{author}{\bibfnamefont{H.~J.} \bibnamefont{Schulz}},
  \bibinfo{journal}{Phys. Rev. B} \textbf{\bibinfo{volume}{52}},
  \bibinfo{pages}{R17040} (\bibinfo{year}{1995}).

\bibitem[{\citenamefont{Maslov and Stone}(1995)}]{Maslov:prb95}
\bibinfo{author}{\bibfnamefont{D.}~\bibnamefont{Maslov}} \bibnamefont{and}
  \bibinfo{author}{\bibfnamefont{M.}~\bibnamefont{Stone}},
  \bibinfo{journal}{Phys. Rev. B} \textbf{\bibinfo{volume}{52}},
  \bibinfo{pages}{R5539} (\bibinfo{year}{1995}).

\bibitem[{\citenamefont{Lal et~al.}(2001)\citenamefont{Lal, Rao, and
  Sen}}]{Lal:prl01}
\bibinfo{author}{\bibfnamefont{S.}~\bibnamefont{Lal}},
  \bibinfo{author}{\bibfnamefont{S.}~\bibnamefont{Rao}}, \bibnamefont{and}
  \bibinfo{author}{\bibfnamefont{D.}~\bibnamefont{Sen}},
  \bibinfo{journal}{Phys. Rev. Lett.} \textbf{\bibinfo{volume}{87}},
  \bibinfo{pages}{026801} (\bibinfo{year}{2001}).

\bibitem[{\citenamefont{Lal et~al.}(2002)\citenamefont{Lal, Rao, and
  Sen}}]{Lal:prb02}
\bibinfo{author}{\bibfnamefont{S.}~\bibnamefont{Lal}},
  \bibinfo{author}{\bibfnamefont{S.}~\bibnamefont{Rao}}, \bibnamefont{and}
  \bibinfo{author}{\bibfnamefont{D.}~\bibnamefont{Sen}},
  \bibinfo{journal}{Phys. Rev. B} \textbf{\bibinfo{volume}{65}},
  \bibinfo{pages}{195304} (\bibinfo{year}{2002}).

\bibitem[{\citenamefont{Safi and Schulz}(1999)}]{Safi:prb99}
\bibinfo{author}{\bibfnamefont{I.}~\bibnamefont{Safi}} \bibnamefont{and}
  \bibinfo{author}{\bibfnamefont{H.~J.} \bibnamefont{Schulz}},
  \bibinfo{journal}{Phys. Rev. B} \textbf{\bibinfo{volume}{59}},
  \bibinfo{pages}{3040} (\bibinfo{year}{1999}).

\bibitem[{\citenamefont{Meden and Schollw\"ok}(2003)}]{Meden:prb03}
\bibinfo{author}{\bibfnamefont{V.}~\bibnamefont{Meden}} \bibnamefont{and}
  \bibinfo{author}{\bibfnamefont{U.}~\bibnamefont{Schollw\"ok}},
  \bibinfo{journal}{Phys. Rev. B} \textbf{\bibinfo{volume}{67}},
  \bibinfo{pages}{193303} (\bibinfo{year}{2003}).

\bibitem[{\citenamefont{Enss et~al.}(2005)\citenamefont{Enss, Meden,
  Andergassen, Barnabe-Theriault, Metzner, and Sch\"onhammer}}]{Enss:prb05}
\bibinfo{author}{\bibfnamefont{T.}~\bibnamefont{Enss}},
  \bibinfo{author}{\bibfnamefont{V.}~\bibnamefont{Meden}},
  \bibinfo{author}{\bibfnamefont{S.}~\bibnamefont{Andergassen}},
  \bibinfo{author}{\bibfnamefont{X.}~\bibnamefont{Barnabe-Theriault}},
  \bibinfo{author}{\bibfnamefont{W.}~\bibnamefont{Metzner}}, \bibnamefont{and}
  \bibinfo{author}{\bibfnamefont{K.}~\bibnamefont{Sch\"onhammer}},
  \bibinfo{journal}{Phys. Rev. B} \textbf{\bibinfo{volume}{71}},
  \bibinfo{pages}{155401} (\bibinfo{year}{2005}).

\bibitem[{Shi({\natexlab{a}})}]{Shimizu}
\bibinfo{note}{A. Shimizu and T. Kato in 'Low-Dimensional Systems ---
  Interactions and Transport Properties' (ed. T. Brandes, Springer, 2000)
  pp.3-22; cond-mat/9911333.}

\bibitem[{Fog()}]{Fogler:prl05}
\bibinfo{note}{M.~M. Fogler and E. Pivovarov, cond-mat/0504502.}

\bibitem[{Kli()}]{Klironomos:cm05}
\bibinfo{note}{A. D. Klironomos, R. R. Ramazashvili, and K. A. Matveev,
  cond-mat/0504118.}

\bibitem[{\citenamefont{Cheianov and
  Zvonarev}(2004{\natexlab{a}})}]{cheianov03}
\bibinfo{author}{\bibfnamefont{V.~V.} \bibnamefont{Cheianov}} \bibnamefont{and}
  \bibinfo{author}{\bibfnamefont{M.~B.} \bibnamefont{Zvonarev}},
  \bibinfo{journal}{Phys. Rev. Lett.} \textbf{\bibinfo{volume}{92}},
  \bibinfo{pages}{176401} (\bibinfo{year}{2004}{\natexlab{a}}).

\bibitem[{\citenamefont{Cheianov and
  Zvonarev}(2004{\natexlab{b}})}]{Cheianov04}
\bibinfo{author}{\bibfnamefont{V.~V.} \bibnamefont{Cheianov}} \bibnamefont{and}
  \bibinfo{author}{\bibfnamefont{M.~B.} \bibnamefont{Zvonarev}},
  \bibinfo{journal}{J. Phys. A: Math. Gen.} \textbf{\bibinfo{volume}{37}},
  \bibinfo{pages}{2261} (\bibinfo{year}{2004}{\natexlab{b}}).

\bibitem[{\citenamefont{Fiete and Balents}(2004)}]{Fiete:prl04}
\bibinfo{author}{\bibfnamefont{G.~A.} \bibnamefont{Fiete}} \bibnamefont{and}
  \bibinfo{author}{\bibfnamefont{L.}~\bibnamefont{Balents}},
  \bibinfo{journal}{Phys. Rev. Lett.} \textbf{\bibinfo{volume}{93}},
  \bibinfo{pages}{226401} (\bibinfo{year}{2004}).

\bibitem[{\citenamefont{Cheianov et~al.}(2005)\citenamefont{Cheianov, Smith,
  and Zvonarev}}]{Cheianov:cm04}
\bibinfo{author}{\bibfnamefont{V.~V.} \bibnamefont{Cheianov}},
  \bibinfo{author}{\bibfnamefont{H.}~\bibnamefont{Smith}}, \bibnamefont{and}
  \bibinfo{author}{\bibfnamefont{M.~B.} \bibnamefont{Zvonarev}},
  \bibinfo{journal}{Phys. Rev. A} \textbf{\bibinfo{volume}{71}},
  \bibinfo{pages}{033610} (\bibinfo{year}{2005}).

\bibitem[{Fie({\natexlab{a}})}]{Fiete:prb05}
\bibinfo{author}{\bibfnamefont{G.~A.} \bibnamefont{Fiete}},
  \bibinfo{author}{\bibfnamefont{J.}~\bibnamefont{Qian}}, 
\bibinfo{author}{\bibfnamefont{Y.}~\bibnamefont{Tserkovnyak}},
\bibnamefont{and}
  \bibinfo{author}{\bibfnamefont{B.~I.} \bibnamefont{Halperin}},
  \bibinfo{journal}{Phys. Rev. B} \textbf{\bibinfo{volume}{72}},
  \bibinfo{pages}{045315} (\bibinfo{year}{2005}).


\bibitem[{\citenamefont{Matveev}(2004{\natexlab{a}})}]{Matveev:prl04}
\bibinfo{author}{\bibfnamefont{K.~A.} \bibnamefont{Matveev}},
  \bibinfo{journal}{Phys. Rev. Lett.} \textbf{\bibinfo{volume}{92}},
  \bibinfo{pages}{106801} (\bibinfo{year}{2004}{\natexlab{a}}).

\bibitem[{\citenamefont{Matveev}(2004{\natexlab{b}})}]{Matveev:prb04}
\bibinfo{author}{\bibfnamefont{K.~A.} \bibnamefont{Matveev}},
  \bibinfo{journal}{Phys. Rev. B} \textbf{\bibinfo{volume}{70}},
  \bibinfo{pages}{245319} (\bibinfo{year}{2004}{\natexlab{b}}).

\bibitem[{\citenamefont{Schulz}(1993)}]{Schulz:prl93}
\bibinfo{author}{\bibfnamefont{H.~J.} \bibnamefont{Schulz}},
  \bibinfo{journal}{Phys. Rev. Lett.} \textbf{\bibinfo{volume}{71}},
  \bibinfo{pages}{1864} (\bibinfo{year}{1993}).

\bibitem[{\citenamefont{Glazman et~al.}(1992)\citenamefont{Glazman, Ruzin, and
  Shklovskii}}]{Glazman:prb92}
\bibinfo{author}{\bibfnamefont{L.~I.} \bibnamefont{Glazman}},
  \bibinfo{author}{\bibfnamefont{I.~M.} \bibnamefont{Ruzin}}, \bibnamefont{and}
  \bibinfo{author}{\bibfnamefont{B.~I.} \bibnamefont{Shklovskii}},
  \bibinfo{journal}{Phys. Rev. B} \textbf{\bibinfo{volume}{45}},
  \bibinfo{pages}{8454} (\bibinfo{year}{1992}).

\bibitem[{Sch()}]{Schulz}
\bibinfo{note}{H. J. Schulz, G. Cuniberti, and P. Pieri, in {\em Field Theories
  for Low-Dimensional Condensed Matter Systems}, ed. by G. Morandi {\em et al.}
  (Springer-Verlag, New York, 2000).}

\bibitem[{\citenamefont{Giamarchi}(2004)}]{Giamarchi}
\bibinfo{author}{\bibfnamefont{T.}~\bibnamefont{Giamarchi}},
  \emph{\bibinfo{title}{Quantum Physics in One Dimension}}
  (\bibinfo{publisher}{Clarendon Press, Oxford}, \bibinfo{year}{2004}).

\bibitem[{\citenamefont{H\"ausler et~al.}(2002)\citenamefont{H\"ausler, Kecke,
  and MacDonald}}]{Hausler:prb02}
\bibinfo{author}{\bibfnamefont{W.}~\bibnamefont{H\"ausler}},
  \bibinfo{author}{\bibfnamefont{L.}~\bibnamefont{Kecke}}, \bibnamefont{and}
  \bibinfo{author}{\bibfnamefont{A.~H.} \bibnamefont{MacDonald}},
  \bibinfo{journal}{Phys. Rev. B} \textbf{\bibinfo{volume}{65}},
  \bibinfo{pages}{085104} (\bibinfo{year}{2002}).

\bibitem[{\citenamefont{Ogata and Shiba}(1990)}]{Ogata:prb90}
\bibinfo{author}{\bibfnamefont{M.}~\bibnamefont{Ogata}} \bibnamefont{and}
  \bibinfo{author}{\bibfnamefont{H.}~\bibnamefont{Shiba}},
  \bibinfo{journal}{Phys. Rev. B} \textbf{\bibinfo{volume}{41}},
  \bibinfo{pages}{2326} (\bibinfo{year}{1990}).

\bibitem[{Shi({\natexlab{b}})}]{Shiba}
\bibinfo{note}{H. Shiba and M. Ogata, in {\em Strongly correlated electron
  systems II}, ed by G. Baskaran, A. E. Ruckenstein, E. Tosatti, and Y. Lu
  (World Scientific, Singapore, 1991), pg. 31.}

\bibitem[{\citenamefont{H\"ausler}(1996)}]{Hausler:zpb96}
\bibinfo{author}{\bibfnamefont{W.}~\bibnamefont{H\"ausler}},
  \bibinfo{journal}{Z. Phys. B} \textbf{\bibinfo{volume}{99}},
  \bibinfo{pages}{551} (\bibinfo{year}{1996}).

\bibitem[{Fie({\natexlab{b}})}]{Fiete:unpublished}
\bibinfo{note}{G. A. Fiete and L. Balents, (unpublished).}

\bibitem[{spi()}]{spin_v_comment}
\bibinfo{note}{An estimate of spin velocities for different potentials can be
  found in Refs.~[\onlinecite{Hausler:prb02}], ~[\onlinecite{Fiete:prb05}], and
  in L. Kecke and W. H\"ausler, Phys. Rev. B {\bf 69}, 085103 (2004) as well as
  M. Fogler, Phys. Rev. B {\bf 71,} 161304(R) (2005).}

\end{thebibliography}

\end{document}